\documentclass[pra,
                preprint,
                amsmath,
                amssymb,
                showpacs,
                superscriptaddress]{revtex4}

\usepackage{graphicx} % Include figure files
\usepackage{dcolumn}  % Align table columns on decimal point
\usepackage{bm}       % bold math
\usepackage{amsfonts}
\usepackage{dsfont}

\begin{document}

\title{Role of electron-electron interference in ultrafast time-resolved imaging of electronic wavepackets}

\author{Gopal Dixit}
\email[]{gopal.dixit@cfel.de}
\affiliation{%
Center for Free-Electron Laser Science, DESY,
            Notkestrasse 85, D-22607 Hamburg, Germany }

\author{Robin Santra}
\email[]{robin.santra@cfel.de}
\affiliation{%
Center for Free-Electron Laser Science, DESY,
            Notkestrasse 85, D-22607 Hamburg, Germany }
\affiliation{%
Department of Physics, University of Hamburg, D-20355 Hamburg, Germany}

\date{\today}

%\pacs{78.70.Ck, 61.05.cf, 82.53.Eb, 87.15.ht}

%%%%%%%%%%%%%%%%% END OF PREAMBLE %%%%%%%%%%%%%%%%

\begin{abstract}
Ultrafast time-resolved x-ray scattering is an emerging approach
to image the dynamical evolution of the electronic charge
distribution during complex chemical and biological processes in
real-space and real-time. Recently, the differences between
semiclassical and quantum-electrodynamical (QED) theory of
light-matter interaction for scattering of ultrashort x-ray pulses
from the electronic wavepacket were formally demonstrated and
visually illustrated by scattering patterns calculated for an
electronic wavepacket in atomic hydrogen [Proc. Natl. Acad. Sci.
U.S.A., {\bf 109}, 11636 (2012)]. In this work, we present a
detailed analysis of time-resolved x-ray scattering from a sample
containing a mixture of non-stationary and stationary electrons
within both the theories. In a many-electron system, the role of
scattering interference between a non-stationary and several
stationary electrons to the total scattering signal is
investigated. In general, QED and semiclassical theory provide
different results for the contribution from the scattering
interference, which depends on the energy resolution of the
detector and the x-ray pulse duration. The present findings are
demonstrated by means of a numerical example of x-ray TRI for an
electronic wavepacket in helium. It is shown that the
time-dependent scattering interference vanishes within
semiclassical theory and the corresponding patterns are dominated
by the scattering contribution from the time-independent
interference, whereas the time-dependent scattering interference
contribution does not vanish in the QED theory and the patterns are
dominated by the scattering contribution from the non-stationary
electron scattering.
\end{abstract}

\maketitle
\section{Introduction}
To fully understand the functionality and dynamic behavior of
molecules, solids and complex biological systems, it is important
to image the motion of electrons in real-time and in real-space.
The motion of atoms within molecules and solids that is associated
with chemical transformations occurs on the femtosecond (1 fs =
10$^{-15}$ s) timescale. The timescale of electronic motion,
responsible for electron-hole dynamics and electron transfer
processes in molecules can be even faster, on the order of
attoseconds (1 as = 10$^{-18}$ s)~\cite{krausz, bucksbaum2007,
corkum2007, smirnova2009, goulielmakis}. The ultimate goal of the
emerging field of time-resolved imaging (TRI) is to visualize
electronic motion on an ultrafast timescale as electrons move in
atoms, complex molecules or solids, as occurring for instance in
photoinduced exciton dynamics, during bond formation and breakage,
conformational changes and charge migration~\cite{breidbach2003,
kuleff2005, remacle, dutoi2011}. With the tremendous advancement
in technology for producing ultraintense and ultrashort x-ray
pulses from novel light sources, it seems possible to obtain
information about ultrafast dynamics of electrons. In extension of
the concept of ``molecular movies'', which track the motion of
atoms on fs timescale~\cite{bergsma1986, bratos2004,
debnarova2006, debnarova2010}, ultrashort, tunable, and
high-energy x-ray pulses from free-electron lasers
(FEL)~\cite{emma2, ishikawa2012}, laser plasmas ~\cite{rousse2001}
and high-harmonic generation~\cite{mckinnie2010, popmintchev2012}
promise to provide ``electronic movies'' that take place on few fs
to as timescale~\cite{vrakking2012}. Recent breakthroughs make it
possible to generate hard \mbox{x-ray} pulses of a few
fs~\cite{emma2, Pile}, and pulse duration of 100~as can in
principle be realized~\cite{Emma1, Zholents}. Utilizing the
remarkable properties of x-rays from FEL, several new insights
have been gained about systems ranging from atoms~\cite{young2010,
rohringer2012}, molecules~\cite{hoener2010, berrah2011},
clusters~\cite{thomas2012}, complex biomolecules~\cite{chapman,
seibert}, to matter in extreme conditions~\cite{vinko2012,
ciricosta2012}.

Since the discovery of x-rays~\cite{rontgen1896}, scattering of
x-rays from matter has been used to unveil the structure of
molecules, solids and biomolecules with atomic-scale spatial
resolution~\cite{Nielsen, bratos2002, lindenberg2005,
hallmann2009, ihee2010ultrafast}. Also, scattering of x-rays from
atoms and molecules has been proposed to gain insight about the
excited electronic states of atomic and molecular
systems~\cite{ben1996, kirrander2012}. In order to image
electronic motion in real-time and real-space with spatial and
temporal resolutions of order 1~{\AA} and 1~fs, respectively, one
can perform scattering of ultrashort \mbox{x-ray} pulses from the
dynamically evolving electronic charge distribution. Pump-probe
experiments are the most direct approach, where first a pump pulse
induces the dynamics and then subsequently a probe pulse
interrogates such induced dynamics. By varying the pump-probe time
delay, one obtains a series of scattering patterns that serve to
image the electronic motion with atomic-scale spatio-temporal
resolution.

Recently, the theory of TRI from a non-stationary electronic
system, using both the semiclassical and the quantum electrodynamical (QED)
treatment of light-matter interaction, has been
developed~\cite{dixit2012}. In a semiclassical theory of
light-matter interaction, matter is treated quantum mechanically
and light is treated classically. In such a situation, the
time-dependent Schr{\"o}dinger equation is solved for the
electrons together with Maxwell's equations for the light. By
solving Maxwell's equations for given charge and current
densities, the expression for the differential scattering
probability (DSP) is obtained, which is a key quantity in x-ray
scattering. According to the semiclassical theory, x-ray TRI would
be expected to provide access to the instantaneous electron
density of the non-stationary electronic system. On the other
hand, in a consistent quantum theory of light-matter interaction,
where both matter and light are treated quantum mechanically,
x-ray TRI encodes the information about spatio-temporal
density-density correlation. Both the theories have been applied
to an electronic wavepacket prepared as a coherent superposition
of eigenstates of atomic hydrogen and it has been shown that the scattering
patterns obtained using both the theories differ drastically from
each other. Moreover, it was shown that the patterns obtained
within QED theory follow the motion of the wavepacket
providing the correct periodicity of the motion, which cannot be
captured by the semiclassical theory. In that case, the notion of
the instantaneous electron density as the key quantity being probed in
x-ray TRI for a sufficiently short pulse completely breaks
down~\cite{dixit2012}. However, x-ray TRI from a single isolated
hydrogen atom is not a realistic scenario. In practice, a sample
contains several electrons and when a tunable pump pulse with
broad energy bandwidth interacts with an $N$-electron system, one
or few electrons participate in the formation of an electronic
wavepacket and other electrons remain stationary. In such a situation,
when an x-ray pulse scatters from a sample containing one or more
non-stationary electrons and several stationary electrons, there
is no way to know whether the scattering has taken place from the non-stationary
electrons or from the stationary electrons and how the
two scattering paths interfere with each other, i.e., interference
between scattering from non-stationary and stationary electrons in the scattering
process. Therefore, at this juncture it is important to analyze
different types of contributions to the total scattering signal.
The total signal can be decomposed into three main parts: first
from stationary electrons, second from non-stationary electrons
and third from the interference between non-stationary and
stationary electrons. In the present work, we will analyze how
these different contributions in an $N$-electron system contribute
to the total scattering signal in both the theories (semiclassical
and QED).

This paper is structured as follows. Section II discusses the
formalism and results for x-ray TRI in the case of many-electron
systems, where only one electron forms an electronic wavepacket and
other electrons serve as stationary reference scatterers in both the
theories. Effects of different parameters such as energy
resolution of the detector, pulse duration and spectral bandwidth
of the x-ray pulse etc. for the electron-electron interference in
the scattering process are discussed in detail. Section III
presents a numerical example of x-ray TRI for an electronic
wavepacket in helium, where one electron forms a coherent
superposition of one-electron eigenstates and the other electron remains
stationary and serves as a reference scatterer. In this particular
situation for helium, the role of the scattering interference is
investigated. Conclusions and future outlook are presented in Sec.
IV.
\section{Theory}
Our investigations are based on the theory for x-ray TRI of
electronic wavepacket motion as developed in
Ref.~\cite{dixit2012}. Our equations are expressed in atomic
units~\cite{szabo1996}.
Under the assumptions that the probe pulse is centered
at the energy of the incident
pulse with very small energy width and the coherence length of the pulse is large in
comparison to the size of the object, the expression for the DSP within the semiclassical theory
is related to the
Fourier transform of the instantaneous electron density,
$\rho_{e}(\mathbf{x}, t)$, as follows~\cite{dixit2012}
\begin{equation}\label{eq1}
\frac{dP}{d\Omega} = \frac{dP_{e}}{d\Omega}
\left|\int d^{3}x\; \rho_{e}(\mathbf{x}, t) ~ e^{i\mathbf{Q \cdot
x}} \right|^{2}.
\end{equation}
Here, $\frac{dP_{e}}{d\Omega}$ is the DSP for a free electron
and $\mathbf{Q}$ is the photon momentum transfer. Here, we have
assumed that the x-ray pulse duration is shorter than the
dynamical timescale of the electronic wavepacket.  According to
Eq.~(\ref{eq1}), the measured scattering pattern provides access
to the instantaneous electron density as a function of the
pump-probe delay time $t$.

Let us consider a scenario for time-resolved scattering of
ultrashort x-rays in order to image the motion of a one-electron
wavepacket in the presence of $N$ stationary electrons. In such
a situation, the $N$ stationary electrons serve as reference
scatterers in the total scattering signal. In this case, using the
language of second quantization~\cite{fetter2003}, the total electronic
wavepacket can be written as
\begin{equation}\label{eq3}
|\Psi(t) \rangle = \hat{c}^{\dagger}_{\varphi(t)} |\Phi_{0}\rangle
= \sum_{a} \alpha_{a} e^{-i \varepsilon_{a} t }
\hat{c}^{\dagger}_{a} |\Phi_{0}  \rangle,
\end{equation}
with
\begin{equation}\label{eq4}
\hat{H} |\Phi_{0} \rangle = \left\{ \sum_{i} \varepsilon_{i}
\right\} |\Phi_{0}  \rangle.
\end{equation}
Here, $\hat{c}^{\dagger}_{p}~(\hat{c}_{p})$ creates (annihilates)
an electron in spin orbital $| \varphi_{p} \rangle$ and
$\varepsilon_{p}$ is the orbital energy corresponding to $|
\varphi_{p} \rangle$, i.e., $\hat{H} | \varphi_{p} \rangle =
\varepsilon_{p} | \varphi_{p} \rangle$.
$\hat{H}$ represents the electronic
Hamiltonian at the mean-field level and $|\Phi_{0}\rangle$ is the unperturbed ground state
of the $N$-electron system
with the electrons filled to the Fermi level. Here and in the
following, indices $p, q, r, s, \ldots$ are used for general spin
orbitals (occupied or unoccupied). Occupied orbitals in
$|\Phi_{0}\rangle$ are presented by indices $i, j, k, l, \ldots$,
whereas unoccupied (virtual) orbitals are symbolized by $a, b, c,
d, \ldots$.

We rewrite the key quantity in Eq.~(\ref{eq1}) in terms of the density operator
\begin{equation}\label{eq8}
\left|\int d^{3}x \rho_{e}(\mathbf{x}, t) ~ e^{i\mathbf{Q \cdot
x}} \right|^{2} = \int d^{3}x \int d^{3}x^{\prime} \langle \Psi(t)
| \hat{n}(\mathbf{x}^{\prime})|\Psi(t) \rangle \langle \Psi(t)
|\hat{n}(\mathbf{x}) |\Psi(t) \rangle e^{i{\mathbf{Q}} \cdot
(\mathbf{x}-\mathbf{x}^{\prime})},
\end{equation}
with
\begin{equation}\label{eq5}
\hat{n}(\mathbf{x}) = \hat{\psi}^{\dagger}(\mathbf{x})
\hat{\psi}(\mathbf{x}) = \sum_{pq}
\varphi^{\dagger}_{p}(\mathbf{x}) \varphi_{q}(\mathbf{x})
\hat{c}^{\dagger}_{p} \hat{c}_{q}.
\end{equation}
Here, $\hat{n}(\mathbf{x})$ is the electron density operator, and
the field operator $\hat{\psi}^{\dagger}(\mathbf{x}) ~
[\hat{\psi}(\mathbf{x})]$ creates (annihilates) an electron at
position $\mathbf{x}$. Using the expression for the wavepacket
as introduced in Eq.~(\ref{eq3}), Eq.~(\ref{eq8}) simplifies as
follows
\begin{subequations}\label{eq9}
\begin{eqnarray}
\left|\int d^{3}x \rho_{e}(\mathbf{x}, t) ~ e^{i\mathbf{Q \cdot
x}} \right|^{2} & = & \left | \sum_{i} \mathcal{L}_{ii} \right|^{2}  \label{eq9a}\\
&& + \sum_{i} \sum_{ab}  \alpha^{*}_{a} \alpha_{b} e^{i(\varepsilon_{a}-\varepsilon_{b})t}
\bigl \{  \mathcal{L}^{*}_{ii} \mathcal{L}_{ab} + \mathcal{L}_{ii} \mathcal{L}^{*}_{ab} \bigr \}  \label{eq9b} \\
& & + \left| \sum_{ab} \alpha^{*}_{a} \alpha_{b}
\mathcal{L}_{ab} e^{i(\varepsilon_{a}-\varepsilon_{b})t}
\right|^{2}, \label{eq9c}
\end{eqnarray}
\end{subequations}
with
\begin{equation}\label{eq6}
\mathcal{L}_{pq} = \int d^{3}x \varphi^{\dagger}_{p}(\mathbf{x})
e^{i{\mathbf{Q}} \cdot \mathbf{x}} \varphi_{q}(\mathbf{x}),
\end{equation}
and
\begin{equation}\label{eq7}
\mathcal{L}^{*}_{pq} = \int d^{3}x
\varphi^{\dagger}_{p}(\mathbf{x}) e^{-i{\mathbf{Q}} \cdot
\mathbf{x}} \varphi_{q}(\mathbf{x}).
\end{equation}
Here, the right-hand side of Eq.~(\ref{eq9a}) provides the
time-independent contributions due to scattering from the
$N$ stationary electrons. The second term as shown
in Eq.~(\ref{eq9b}) is due to the scattering interference between
the $N$ stationary electrons and the non-stationary electron. The
last time-dependent term in Eq.~(\ref{eq9c}) is solely due to
scattering from the non-stationary electron.

On the other hand, in the full quantum theory of x-ray TRI, both matter
and x-ray pulse are treated quantum mechanically and first-order
time-dependent perturbation theory is employed for the
interaction between matter and x rays. Here we assume that
the probe pulse has a small bandwidth and a small angular spread so
that the pixel assignment is well defined in the momentum space,
the coherence length of the pulse is large in comparison to the
size of the object, and the pulse duration should be sufficiently short to freeze the
dynamics of the electronic wavepacket. Under these assumptions, the resulting
expression for the DSP from a coherent, Gaussian \mbox{x-ray}
pulse is~\cite{dixit2012}
\begin{eqnarray}\label{eq2}
\frac{dP}{d\Omega} & = &  \frac{dP_{e}}{d\Omega} \int_{0}^{\infty}
d\omega_{\mathbf{k}_{s}} ~ W_{\Delta E}
({\omega_{\mathbf{k}_{s}}}) ~
\frac{\omega_{\mathbf{k}_{s}}}{\omega_{\mathbf{k}_{in}}} ~
\int_{-\infty}^{\infty}
\frac{d\tau}{2\pi} ~C(\tau) ~ e^{- i(\omega_{\mathbf{k}_{s}}-
\omega_{\mathbf{k}_{in}}) \tau }\nonumber \\
& & \times \int d^{3}x \int d^{3}x^{\prime} ~ \left \langle \Psi
\left( t +\frac{\tau}{2} \right) \Biggl| ~
\hat{n}\left(\mathbf{x}^{\prime} \right)~
e^{-i\hat{H}\tau}~\hat{n}\left(\mathbf{x}\right)~ \Biggr| \Psi
\left( t-\frac{\tau}{2}\right) \right \rangle e^{i{\mathbf{Q}}
\cdot (\mathbf{x}-\mathbf{x^\prime})}.
\end{eqnarray}
Here, $C(\tau) = \exp({-\frac{2\ln{2}\;\tau^{2}}{\tau_{l}^{2}}})$
is a function of the pulse duration $\tau_{l}$.
$\omega_{\mathbf{k}_{in}}$ and $\omega_{\mathbf{k}_{s}}$ refer to the energy of the incident and scattered
photon, respectively, while $W_{\Delta E}({\omega_{\mathbf{k}_{s}}})$ is a
spectral window function centered at $\omega_{\mathbf{k}_{in}}$
with a width $\Delta E$. $W_{\Delta E}({\omega_{\mathbf{k}_{s}}})$
models the range of energies of the scattered photons accepted by
the detector.

In the case of QED theory, an energy-resolved scattering process is
considered for x-ray TRI.
Therefore, any
inelastic (Compton) scattering contributions due to excitations from the
$N$ stationary electrons can be easily distinguished by
utilizing the energy-resolving detector, if we assume that $\Delta E$ is small in comparison to the characteristic excitation energies of
$|\Phi_{0} \rangle$. Hence,
excitations from the $N$ stationary electrons are not considered
in the following.
Similarly, on using the expression for the
wavepacket as introduced in Eq.~(\ref{eq3}), the key expression of
Eq.~(\ref{eq2}) is simplified as follows
\begin{subequations}\label{eq10}
\begin{eqnarray}
\lefteqn{ \int d^{3}x \int d^{3}x^{\prime} ~ \left \langle \Psi
\left( t +\frac{\tau}{2} \right) \Biggl| ~
\hat{n}\left(\mathbf{x}^{\prime} \right)~
e^{-i\hat{H}\tau}~\hat{n}\left(\mathbf{x}\right)~ \Biggr| \Psi
\left( t-\frac{\tau}{2}\right) \right \rangle
e^{i{\mathbf{Q}} \cdot (\mathbf{x}-\mathbf{x^\prime})}} \nonumber \\
& = & \left| \sum_{i} \mathcal{L}_{ii} \right |^{2} \label{eq10a} \\
& & + \sum_{i} \sum_{ab}  \alpha^{*}_{a} \alpha_{b} e^{i(\varepsilon_{a}-\varepsilon_{b})t}
\bigl \{ \mathcal{L}^{*}_{ii} \mathcal{L}_{ab} e^{-i(\varepsilon_{a}-\varepsilon_{b})\frac{\tau}{2}} +
\mathcal{L}_{ii} \mathcal{L}^{*}_{ab} e^{i(\varepsilon_{a}-\varepsilon_{b})\frac{\tau}{2}} \bigr \} \label{eq10b} \\
& & +  \sum_{a b} \sum_{c} \alpha^{*}_{a} \alpha_{b}
\mathcal{L}^{*}_{ac} \mathcal{L}_{cb}
e^{i(\varepsilon_{a}-\varepsilon_{b})t}
e^{i(\varepsilon_{a}+\varepsilon_{b}-2\varepsilon_{c})\frac{\tau}{2}}.
\label{eq10c}
\end{eqnarray}
\end{subequations}
Here, the first term, Eq.~(\ref{eq10a}), provides the
time-independent contribution due to scattering from the
$N$ stationary electrons, which is identical to
Eq.~(\ref{eq9a}). The second term in
Eq.~(\ref{eq10b}) is due to the scattering interference between
the $N$ stationary electrons and the non-stationary electron,
which seems different to the one shown in Eq.~(\ref{eq9b}). The
last time-dependent term, Eq.~(\ref{eq10c}), is again solely due
to scattering from the non-stationary electron. The scattering
contributions from the non-stationary electron, Eqs.~(\ref{eq9c})
and (\ref{eq10c}), are not identical and provide completely
different information about the electronic motion as shown
in the case of a one-electron wavepacket in atomic
hydrogen~\cite{dixit2012}. It is evident from Eqs.~(\ref{eq9})
and (\ref{eq10}), that the first term in both the theories,
Eqs.~(\ref{eq9a}) and (\ref{eq10a}), provides identical scattering
contributions to the total signal. In many systems of interest, the number of
stationary electrons is large and therefore, the
time-independent terms contribute a strong static background in
the total signal. Due to the large number of the stationary electrons
and one or few non-stationary electrons, the dominating
time-dependent contributions in the total scattering signal are
due to the scattering interference between stationary and
non-stationary electrons
(unless one is considering $\mathbf{Q}$ for which $\sum_{i} \mathcal{L}_{ii}$ is small).
Therefore, it is crucial to analyze the
scattering interference term, Eqs.~(\ref{eq9b}) and (\ref{eq10b}),
in both the theories.

On substituting Eqs.~(\ref{eq9b}) and (\ref{eq10b}) in
Eqs.~(\ref{eq1}) and (\ref{eq2}), respectively, the expression
for the scattering interference contribution to the DSP,
$\frac{dP_{\textrm{int}}}{d\Omega}$, can be written as
\begin{equation}\label{eq12}
\frac{dP_{\textrm{int}}}{d\Omega} = \frac{dP_{e}}{d\Omega}
\sum_{i} \sum_{ab} \alpha^{*}_{a} \alpha_{b}
e^{i(\varepsilon_{a}-\varepsilon_{b})t} \bigl \{
\mathcal{L}^{*}_{ii} \mathcal{L}_{ab} + \mathcal{L}_{ii}
\mathcal{L}^{*}_{ab} \bigr \}
\end{equation}
in the semiclassical theory, and in the QED formalism can be
expressed as
\begin{eqnarray}\label{eq11}
\frac{dP_{\textrm{int}}}{d\Omega} & = & \frac{dP_{e}}{d\Omega}
\int_{0}^{\infty}
d\omega_{\mathbf{k}_{s}} ~ W_{\Delta E}
({\omega_{\mathbf{k}_{s}}})
~\frac{\omega_{\mathbf{k}_{s}}}{\omega_{\mathbf{k}_{in}}}
\frac{\tau_{l}}{\sqrt{8 \pi \textrm{ln 2}}}
 \sum_{i} \sum_{ab} \alpha^{*}_{a} \alpha_{b} e^{i(\varepsilon_{a}-\varepsilon_{b})t} \nonumber \\
& & \times \left \{\mathcal{L}^{*}_{ii} \mathcal{L}_{ab}
e^{-\frac{\tau_{l}^{2}}{8\textrm{ln
2}}\left(\omega_{\mathbf{k}_{s}}-
\omega_{\mathbf{k}_{in}}+(\frac{\varepsilon_{a}-\varepsilon_{b}}{2})\right)^{2}}
+ \mathcal{L}_{ii} \mathcal{L}^{*}_{ab}
e^{-\frac{\tau_{l}^{2}}{8\textrm{ln
2}}\left(\omega_{\mathbf{k}_{s}}-
\omega_{\mathbf{k}_{in}}-(\frac{\varepsilon_{a}-\varepsilon_{b}}{2})\right)^{2}}
\right\}.
\end{eqnarray}
It is important to analyze several aspects of both expressions,
Eqs.~(\ref{eq12}) and (\ref{eq11}), under different circumstances
and properties of the probe pulse.
\begin{itemize}
\item
The Gaussian distributions  in Eq.~(\ref{eq11}) are centered
at positions $\omega_{\mathbf{k}_{in}}+(\frac{\varepsilon_{a}-\varepsilon_{b}}{2})$
and $\omega_{\mathbf{k}_{in}}-(\frac{\varepsilon_{a}-\varepsilon_{b}}{2})$, and the width of the distributions is
determined by $\tau_{l}$.  Therefore, if $\Delta E$ is larger than the separation between the two distributions and
the detector has a lower
energy resolution than the energy spectral bandwidth of the x-ray
pulse, then all
the scattered photons are detected by the
detector and $W_{\Delta E}$ is constant in the energy range contributing to the integral in Eq.~(\ref{eq11}).
Hence, on performing the energy integral,
Eq.~(\ref{eq11}) reduces to Eq.~(\ref{eq12}). However, when
the energy resolution of the detector is poor, it is very
difficult to assign a unique pixel in $Q$-space to the scattered photon in the
detector and it may no longer be possible to filter out Compton scattering from the $N$ stationary electrons.
\end{itemize}
\begin{itemize}
\item If the probe pulse is very short in comparison to the dynamical timescale of the electronic motion,
the $\tau$ dependent exponent in Eq.~(\ref{eq10b})  will reduce to
unity. Therefore, if $\Delta E$ is smaller than the energy spectral width of the pulse  
and centered at the incident energy such that
$\omega_{\mathbf{k}_{s}} \approx \omega_{\mathbf{k}_{in}}$,
the energy integral, Eq.~(\ref{eq11}) reduces to Eq.~(\ref{eq12})
and the scattering interference contributions from both the theories
will be identical.  However,
an ultrashort pulse corresponds to a large energy spectral width due to
the energy-time uncertainty relation. Therefore, the unavoidable energy
bandwidth of the pulse causes an uncertainty in the momentum
distribution of the incoming photon. In such a situation
uniqueness of a pixel in $Q$-space is lost.
\end{itemize}
\begin{itemize}
\item In general, however, the probe pulse is not very short in comparison to the dynamical timescale
of the electronic motion. In such situations the $\tau$ dependent exponent in Eq.~(\ref{eq10b})  will not reduce to unity
and both expressions for the interference  will provide different contributions to the total signal.
\end{itemize}

In the following subsection, we present an example of
one-electron wavepacket motion in helium. In this example, we
apply both the approaches to compute time-resolved scattering
patterns and analyze the contributions from the scattering interference to the
total scattering signal.
\section{Numerical Results and Discussion}
\begin{figure*}
\includegraphics[width=15cm]{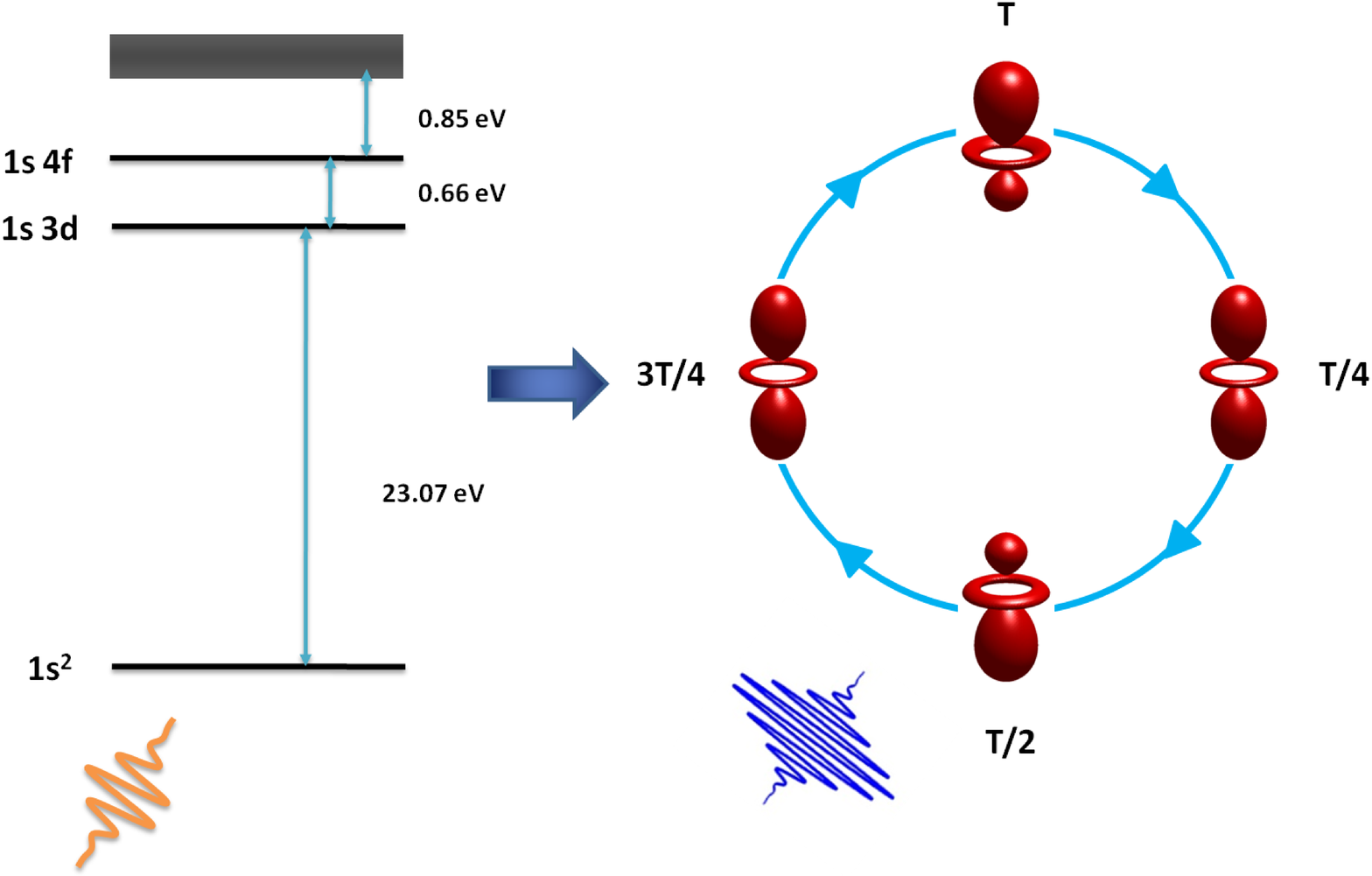}
\caption{A schematic scenario for probing an electronic
wavepacket motion using ultrafast time-resolved x-ray scattering in helium
atom. A pump pulse with broad energy bandwidth (indicated in
orange) excites one of the electrons from the ground state and forms a
coherent superposition of the 1s3d and 1s4f eigenstates with equal
population. An isosurface of the electronic charge distribution of
the wavepacket is shown (indicated in red), which undergoes periodic
oscillation with oscillation period T = 6.25 fs. The
dynamically evolving electronic charge distribution is probed by an ultrafast
x-ray pulse (indicated in blue). By varying the pump-probe time
delay, one obtains a series of scattering patterns that serve to
image the electronic motion with atomic-scale spatio-temporal
resolution.} \label{fig1}
\end{figure*}

A schematic scenario for probing an electronic wavepacket motion
in helium is shown in Fig.~\ref{fig1}. The ground state
configuration for both the electrons is 1s$^2$. The ionization
potential of the first electron is 24.59 eV and 54.42 eV for the
second electron~\cite{nist}. A pump pulse with broad energy
bandwidth excites one of the electrons from the ground state
configuration and prepares a coherent superposition of the 1s3d and
1s4f configurations with the projection of orbital angular
momentum being equal to zero. The energy difference between the 1s$^2$
and 1s3d configurations is 23.07 eV and between the 1s3d and 1s4f
configurations is 0.66 eV (see Fig.~\ref{fig1}). Therefore, the
dynamical timescale of the electronic wavepacket motion, which is
inversely related to the energy spacing between the eigenstates
participating in the wavepacket, is 6.25 fs. The
spatial extension of the wavepacket  is 14--17 \AA~along the $z$
axis and 7.5--9 \AA~along the $x$ and $y$ axes. It is known that
for orbital angular momentum quantum number equal to or larger than
two, the quantum defect is almost zero. Therefore, the electron in
the $1s$ orbital sees no shielding of the nuclear charge and the electron
in the superposition of $3d$ and $4f$ orbitals sees complete
shielding of the nuclear charge. Therefore, the wavefunction for the
stationary electron in the $1s$ orbital can be expressed in terms of
the hydrogenic wavefunction with nuclear charge Z = 2. On the other
hand, the wavefunction for the non-stationary electron in $3d$ and
$4f$ orbitals as well as other higher-lying orbitals
can be written in terms of the corresponding
hydrogenic wavefunction with Z = 1. This type of procedure for
treating the two-electron problem has already been used in the
past and was successfully applied to describe different types of
physical processes~\cite{bethe1977}.

In order to compute time-dependent scattering patterns of the
electronic wavepacket (cf. Fig.~\ref{fig1}) as a function of the
delay time in helium, we employ both
the semiclassical and the QED approaches, i.e., Eqs.~(\ref{eq1}) and
(\ref{eq2}). Since the non-stationary and stationary electrons are
energetically distinguishable (energy difference is around 23 eV)
and an energy-resolved scattering process is considered with energy
resolution at least equal to the unavoidable spectral bandwidth of the
probe pulse, any excitation from the stationary
electron can be easily filtered out
and therefore is not
considered in the present case. The scattering operator $e^{i{\mathbf{Q}} \cdot \mathbf{x}}$ is
expanded in terms of the spherical Bessel functions, $j_{l}({Q}\,
{r})$, and spherical harmonics $Y^{l}_{m}(\theta, \phi)$ as
\begin{equation}\label{eq201}
e^{i{\mathbf{Q}} \cdot \mathbf{x}}  =  4 \pi \sum_{l, m} i^{l} ~
j_{l}({Q}\, {r}) ~  {Y^{l}_{m}}^{*}(\alpha, \beta) ~
Y^{l}_{m}(\theta, \phi),
\end{equation}
where $r = |\mathbf{x}|$. After introducing hydrogenic
wavefunctions, the expression in
Eq.~(\ref{eq2}) factorizes into radial and angular parts. The
angular part is given by
\begin{eqnarray}\label{eq202}
\int_{0}^{2 \pi} \int_{0}^{\pi} Y^{l_{1}}_{m_{1}}(\theta, \phi)
Y^{l_{2}}_{m_{2}}(\theta, \phi) Y^{l_{3}}_{m_{3}}(\theta, \phi)
\sin\theta~ d\theta d\phi
& = & \sqrt \frac {(2l_{1}+1)(2l_{2}+1)(2l_{3}+1)}{4 \pi}  \nonumber \\
&  & \times  \left (\begin{array}{ccc}
l_{1} & l_{2} & l_{3} \\
0 & 0 & 0
\end{array} \right)
\left (\begin{array}{ccc}
l_{1} & l_{2} & l_{3} \\
m_{1} & m_{2} & m_{3}
\end{array} \right),
\end{eqnarray}
whereas the radial part is numerically integrated. To calculate
the patterns as a function of the delay times, we used a
Gaussian pulse of duration 1 fs with 4 keV incoming photon energy,
and assumed a Gaussian photon energy detection window of width
$\Delta E=$~0.5~eV for the detector.  The patterns are calculated
for ${Q}_{\mathrm{max}} = 2$ \AA$^{-1}$ corresponding to a 3.14
\AA~spatial resolution and to a detection angle of scattered
photons of up to 60$^{\circ}$. In order to compute the scattering
contribution from the non-stationary electron to the total
scattering signal, all transitions induced during the scattering
process within the energy detection window are
computed~\cite{dixit2012}. Therefore, transition amplitudes from
the eigenstates involved in the electronic wavepacket to all
the electronic states within the detection range of $\Delta E$ are
computed, which includes all types of multipole transitions
allowed by the conservation of angular momentum and amounts to
22000 transition amplitudes. Also, the scattering patterns using
semiclassical theory are calculated by convolution of the square
of the Fourier transform of $\rho_{e}(\mathbf{x}, t)$ with an x-ray
pulse of duration 1 fs.

\begin{figure*}
\includegraphics[width=17cm]{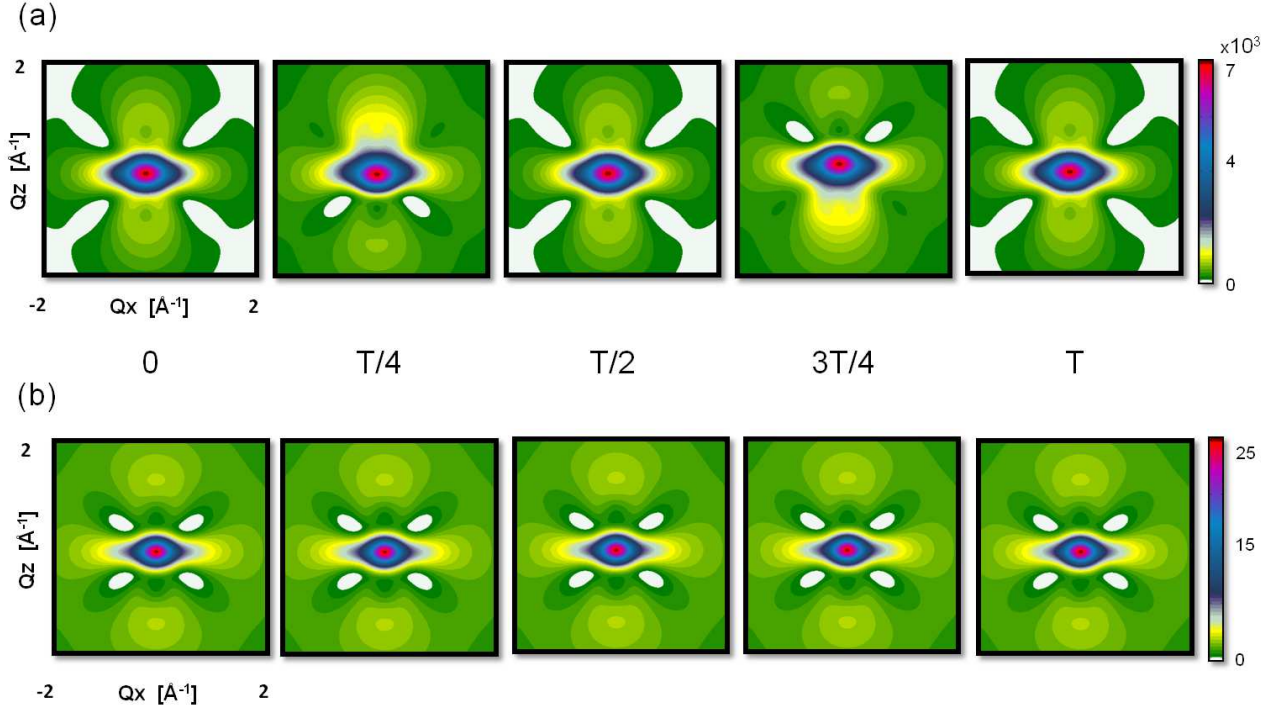}
\caption{Scattering patterns in the $Q{_x}$ -
$Q{_z}$ plane ($Q{_y}$ = 0) of helium (cf. Fig. 1).
Scattering patterns obtained using (a) QED theory, i.e., Eq.~(\ref{eq2}) and (b)
semiclassical theory, i.e., Eq.~(\ref{eq1}) at pump-probe delay times 0, T/4, T/2, 3T/4, and
T, where the oscillation period of the electronic wavepacket is T
= 6.25 fs.} \label{fig5}
\end{figure*}

Scattering patterns in the $Q_x$ - $Q_z$ plane ($Q_y$ = 0) as a
function of the delay time at times 0, T/4, T/2, 3T/4, and T are
depicted in Fig.~\ref{fig5}. The time-dependent patterns shown in
Fig.~\ref{fig5}(a) are computed within QED theory using
Eq.~(\ref{eq2}), whereas in Fig.~\ref{fig5}(b) are computed within
semiclassical theory using Eq.~(\ref{eq1}). It is evident from
Fig.~\ref{fig5}(a) that the patterns undergo spatial oscillation
along $Q_z$ in momentum space and reflect the motion of
the wavepacket along $z$ in real space (see
Fig.~\ref{fig1}). 
One of the striking features of the patterns shown in Fig.~\ref{fig5}(a) is that when the charge distributions
are symmetric, and corresponding patterns are asymmetric and vice versa, which can be understood as follows.
The charge distributions are identical at delay times T/4 and 3T/4, as may be seen in Fig.~\ref{fig1},
while the electron clouds move in opposite directions at the two times. At time T/4, the flow of the
electron cloud is downwards, whereas at time 3T/4 the flow is upwards. This is reflected
by their corresponding patterns. Therefore, the patterns calculated within the QED theory capture the dynamics
of the {\em momentum distribution} of the wavepacket. As a consequence, the apparent motions of the charge 
distributions and of the scattering patterns are shifted by 90$^{\circ}$.
On the other hand, the patterns shown in Fig.~\ref{fig5}(b) do not
change significantly as a function of the delay time.

\begin{figure*}
\includegraphics[width=17cm]{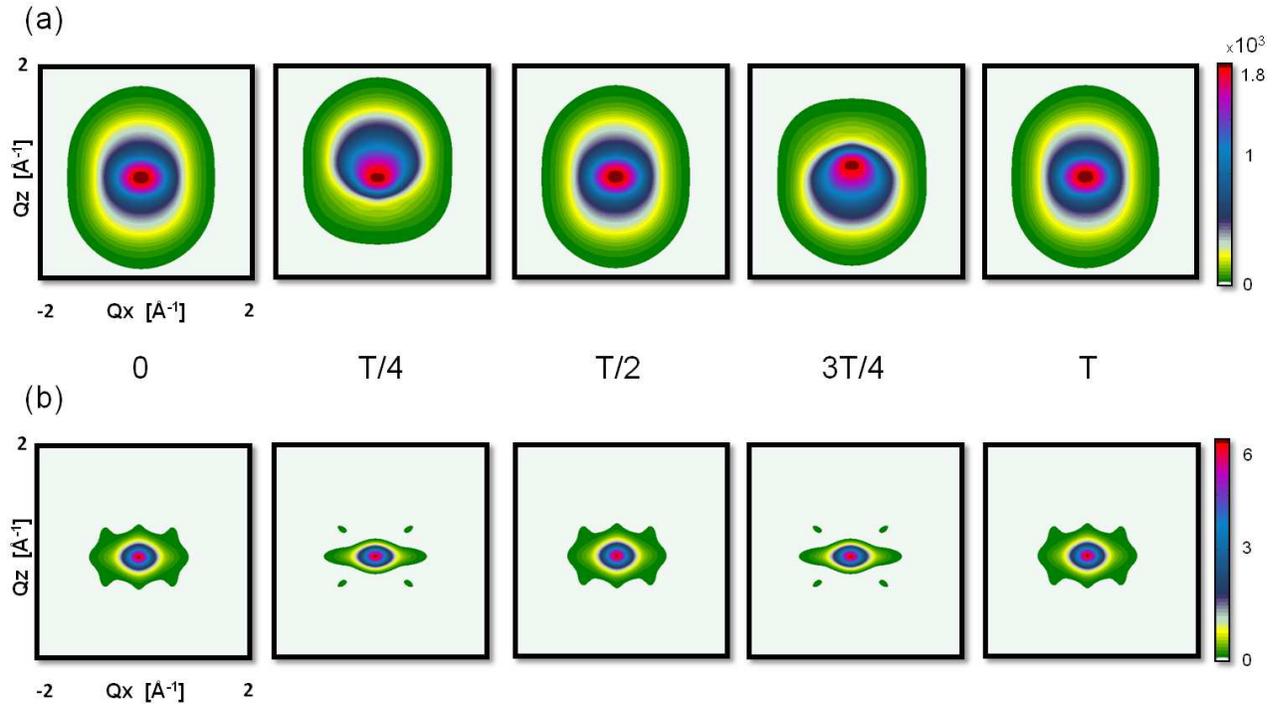}
\caption{Scattering contributions from the
non-stationary electron to the total scattering patterns in the
$Q{_x}$ - $Q{_z}$ plane ($Q{_y}$ = 0) in helium. Scattering
patterns obtained using (a) Eq.~(\ref{eq10c}) and (b)
Eq.~(\ref{eq9c}) at pump-probe delay times 0, T/4, T/2, 3T/4, and
T, where the oscillation period of the electronic wavepacket is T
= 6.25 fs.} \label{fig2}
\end{figure*}

In order to understand the scattering from the non-stationary
electron, the scattering patterns corresponding to the
non-stationary electron are shown in Fig.~\ref{fig2} in the $Q_x$
- $Q_z$ plane ($Q_y$ = 0) as a function of the delay time. The
patterns shown in Fig.~\ref{fig2}(a) are obtained using
Eq.~(\ref{eq10c}) and in Fig.~\ref{fig2}(b) are obtained using
Eq.~(\ref{eq9c}). It is evident from Fig.~\ref{fig2}(a) that the
scattering patterns undergo oscillations as a function of the
delay time. The scattering patterns shown in
Fig.~\ref{fig2}(b) are localized in the low $Q$ region,
which reflects the spatial extension of the electronic charge
distribution of the wavepacket and also undergo changes as a function of the delay
time, but do not display oscillations as the charge distribution
oscillates.  Hence, patterns obtained using
semiclassical theory provide half of the actual period of the
motion as the patterns start repeating themselves in half of the
actual time of the motion.

We now investigate the contribution from the scattering interference
between the stationary and non-stationary electrons to the total
scattering signal in both the theories.
On re-writing Eqs.~(\ref{eq6}) and (\ref{eq7}) in the case of
helium, we find that
\begin{equation}\label{eq61}
\mathcal{L}_{1s1s} = \int d^{3}x \varphi^{\dagger}_{1s
}(\mathbf{x}) e^{i{\mathbf{Q}} \cdot \mathbf{x}} \varphi_{1s
}(\mathbf{x})
\end{equation}
is a purely real number, as is evident from Eqs.~(\ref{eq201}) and (\ref{eq202}),
whereas
\begin{equation}\label{eq71}
\mathcal{L}_{3d4f} = \int d^{3}x
\varphi^{\dagger}_{3d}(\mathbf{x}) e^{i{\mathbf{Q}} \cdot
\mathbf{x}} \varphi_{4f}(\mathbf{x})
\end{equation}
is a purely imaginary number, which also follows from Eqs.~(\ref{eq201}) and (\ref{eq202}).
Therefore,
\begin{equation}\label{eq62}
\mathcal{L}_{1s1s} = \mathcal{L}^{*}_{1s1s} ,
\end{equation} and
\begin{equation}\label{eq72}
\mathcal{L}_{3d4f} = -\mathcal{L}^{*}_{3d4f}.
\end{equation}
In a similar way we can write $\mathcal{L}_{3d3d} = \mathcal{L}^{*}_{3d3d}$, and
$\mathcal{L}_{4f4f} = \mathcal{L}^{*}_{4f4f}$, which are both real.
On substituting the contributions of the $\mathcal{L}$'s in
Eq.~(\ref{eq9b}), the contribution from the scattering interference within
the semiclassical theory can be written as
\begin{equation}\label{eq631}
\sum_{a = 3d,4f} ~ \sum_{b = 3d,4f} \alpha^{*}_{a} \alpha_{b}
e^{i(\varepsilon_{a}-\varepsilon_{b})t} \bigl \{
\mathcal{L}^{*}_{1s1s} \mathcal{L}_{ab} + \mathcal{L}_{1s1s}
\mathcal{L}^{*}_{ab} \bigr \},
\end{equation}
which can be further decomposed into two parts: a time-dependent
scattering interference contribution
\begin{subequations}\label{eq632}
\begin{eqnarray}
\lefteqn{ \alpha^{*}_{3d} \alpha_{4f}
e^{i(\varepsilon_{3d}-\varepsilon_{4f})t} \bigl \{
\mathcal{L}^{*}_{1s1s} \mathcal{L}_{3d4f} + \mathcal{L}_{1s1s}
\mathcal{L}^{*}_{3d4f} \bigr \} + \alpha^{*}_{4f} \alpha_{3d}
e^{i(\varepsilon_{4f}-\varepsilon_{3d})t}  \bigl \{
\mathcal{L}^{*}_{1s1s} \mathcal{L}_{4f3d} + \mathcal{L}_{1s1s}
\mathcal{L}^{*}_{4f3d} \bigr \} } \nonumber \\
& =& \alpha^{*}_{3d} \alpha_{4f}
e^{i(\varepsilon_{3d}-\varepsilon_{4f})t}
 \mathcal{L}_{1s1s}  \bigl \{ \mathcal{L}_{3d4f} -  \mathcal{L}_{3d4f} \bigr
 \} +  \alpha^{*}_{4f} \alpha_{3d}
e^{i(\varepsilon_{4f}-\varepsilon_{3d})t}
 \mathcal{L}_{1s1s}  \bigl \{ \mathcal{L}_{4f3d} -  \mathcal{L}_{4f3d} \bigr
 \} = 0, \nonumber
\end{eqnarray}
\end{subequations}
which is zero, and a time-independent scattering interference
contribution
\begin{equation}\label{eq633}
 2 \mathcal{L}_{1s1s} [
|\alpha_{3d}|^{2}
  \mathcal{L}_{3d3d}  + |\alpha_{4f}|^{2}
 \mathcal{L}_{4f4f}].
\end{equation}
Similarly on substituting the contributions of the $\mathcal{L}$'s in
Eq.~(\ref{eq10b}), the contribution from the scattering
interference within the QED theory can be written as the sum of two parts:
a time-dependent scattering interference contribution
\begin{subequations}\label{eq73}
\begin{eqnarray}
\lefteqn{  \alpha^{*}_{3d} \alpha_{4f}
e^{i(\varepsilon_{3d}-\varepsilon_{4f})t} \mathcal{L}_{1s1s}
\mathcal{L}_{3d4f} \bigl \{
e^{-i(\varepsilon_{3d}-\varepsilon_{4f})\frac{\tau}{2}} -
e^{i(\varepsilon_{3d}-\varepsilon_{4f})\frac{\tau}{2}} \bigr \} } \label{eq73a} \\
& & + \alpha^{*}_{4f} \alpha_{3d}
e^{i(\varepsilon_{4f}-\varepsilon_{3d})t} \mathcal{L}_{1s1s}
\mathcal{L}_{4f3d} \bigl \{
e^{-i(\varepsilon_{4f}-\varepsilon_{3d})\frac{\tau}{2}} -
e^{i(\varepsilon_{4f}-\varepsilon_{3d})\frac{\tau}{2}} \bigr \},
\label{eq73b}
\end{eqnarray}
\end{subequations}
which is non-zero, and a time-independent scattering interference
contribution
\begin{equation}\label{eq74}
 2 \mathcal{L}_{1s1s} [
|\alpha_{3d}|^{2}
  \mathcal{L}_{3d3d}  + |\alpha_{4f}|^{2}
 \mathcal{L}_{4f4f}],
\end{equation}
which is identical to the one obtained within semiclassical theory.
We rewrite the time-dependent interference contribution, Eq~(\ref{eq73}), to
the leading nonvanishing order in $\tau$ as
\begin{equation}\label{eq75}
i\mathcal{L}_{1s1s} (\varepsilon_{4f}-\varepsilon_{3d})\tau \bigl [\alpha^{*}_{3d} \alpha_{4f} \mathcal{L}_{3d4f}
e^{i(\varepsilon_{3d}-\varepsilon_{4f})t}-
\alpha^{*}_{4f} \alpha_{3d} \mathcal{L}_{4f3d} e^{i(\varepsilon_{4f}-\varepsilon_{3d})t} \bigr]  \neq 0.
\end{equation}
Therefore, for the particular combination of orbitals involved in
the one-electron wavepacket and the orbital corresponding to
the stationary electron in helium (see Fig.~\ref{fig1}), the
time-dependent scattering interference between non-stationary and
stationary orbitals is zero within the semiclassical theory and
negligibly small in comparison  to the total scattering signal in
the QED theory. However, the time-independent scattering interference
between the orbitals, i.e., the interference between $1s$ and $3d$
orbitals and $1s$ and $4f$ orbitals, contributes equally to the
total scattering signal in both the theories, as is evident from
Eqs.~(\ref{eq633}) and  ~(\ref{eq74}). The reason why
the time-dependent scattering interference
contribution to the total signal is so small in QED theory can be understood
as follows: For probing the ultrafast motion, one has to satisfy
$\Delta \varepsilon ~\tau \ll 1$, where $\Delta \varepsilon = \varepsilon_{4f}-\varepsilon_{3d}$ is the
characteristic energy scale of the electronic wavepacket, i.e., the
pulse duration of the probe pulse should be much smaller than the
characteristic  timescale of the motion.
Thus, the contribution from Eq.~(\ref{eq75}), which is proportional to $\Delta \varepsilon ~\tau$, is suppressed.

On comparing the scattering contribution from the non-stationary
electron with respect to total scattering signal, i.e., comparing
Figs.~\ref{fig2}(a) and ~\ref{fig5}(a), one can easily distinguish the
scattering contributions from the stationary electron and
the interferences to the total
scattering patterns in the QED theory. In Fig.~\ref{fig5}(a), the
broadening of the scattering signal in the high $Q$ region is the
reflection of the contribution from the time-independent
scattering interference between orbitals, whereas the wing type
structures along the diagonal, which change as a function of the delay
time, are a reflection of the contribution from the time-dependent
scattering interference between orbitals, which is weak in
comparison to the total scattering signal. On the other hand, on
comparing the patterns shown in Figs.~\ref{fig5}(b) and
~\ref{fig2}(b), one can only observe the broadening of scattering
patterns, which reflects the contribution from the time-independent
scattering interference between orbitals.

Therefore, the patterns within the semiclassical theory
are dominated by the scattering contribution from the time-independent
interference between orbitals, and are not changing significantly
as a function of the delay time. In contrast, the patterns within
the QED theory are dominated by the scattering contributions from the non-stationary electron due
to Compton scattering within the finite energy detection range of the detector.
\section{Conclusions}
This work is devoted to understanding ultrafast time-resolved x-ray
scattering from a sample containing a mixture of a non-stationary
electron in the form of a one-electron wavepacket and one or more
stationary electrons using the semiclassical theory and the QED theory of
light-matter interaction. The contributions of the scattering
interference between the non-stationary and the stationary
electrons to the total time-dependent scattering signal are
investigated in both the theories. Our investigations are based on
the recent theory for time-resolved x-ray scattering to image the
electronic wavepacket motion~\cite{dixit2012}. First, we
investigated different scattering contributions to the total
scattering signal in both the theories and showed that the
total signal can be decomposed into three main scattering
contributions: first from the stationary electrons, second from
the non-stationary electron and third from the interference between
stationary and non-stationary electrons. In both the theories, the
scattering contributions from the stationary electrons to the
signal are identical, whereas scattering contributions from the
non-stationary electron are completely different.
In the QED theory, the scattering contributions from the interference
depend on the energy resolution of the detector and the x-ray
pulse duration. Therefore, in case of negligible energy resolution
or extremely short pulses, QED theory provides identical
contributions for the scattering interference as one obtaines in the semiclassical theory.
On the other hand, if the pulse duration is not very short in comparison to the
dynamical timescale of the motion and if the energy
resolution is sufficiently high, the scattering interference in
the QED theory does not provide identical result to the one obtaines in the semiclassical theory.
It is important to note that
most organic molecules and proteins contain mainly
hydrogen, carbon, nitrogen and oxygen atoms and in such cases the
scattering signal is dominated by the scattering from carbon,
nitrogen and oxygen atoms. In these atoms, only two electrons are
deeply bound core electrons, whereas other electrons are loosely
bound valence electrons. When a pump pulse with broad bandwidth
initiates excitation in such atoms it might be possible that more
than one electron participates in the formation of the electronic
wavepacket. In such situations, the time-dependent scattering
signal would be dominated by the scattering contributions from
the non-stationary electrons rather than the stationary and the
interference contributions.

Both the theories for light-matter interaction are illustrated by
means of calculating the time-dependent scattering patterns for a one-electron
wavepacket in helium. In helium, the pump pulse
excites one of the electrons from the ground state and prepares an
electronic wavepacket as a coherent superposition of the 1s3d and 1s4f
eigenstates. The scattering patterns are computed for the
non-stationary electron in the presence of a stationary electron.
The time-dependent interference between the stationary and non-stationary electrons within the
semiclassical theory is zero, and it is quite small in
comparison to the total scattering signal in the QED theory. However,
the time-independent interference between the stationary and non-stationary electrons contributes
identically to the total signal in both the theories. The patterns
are dominated by the scattering contribution from the
time-independent interference within the
semiclassical theory, whereas the patterns are dominated by the
scattering contributions from the non-stationary electron
due to Compton scattering within the QED
theory. Henceforth, the dynamical features of the patterns cannot
be captured within semiclassical theory. We expect that our present analysis
of TRI using ultrafast x-ray scattering will find several
important applications for exploring ultrafast dynamics in nature.
With the recent advent of novel light sources, we also believe
that our findings will shed light on ultrafast electronic motion,
for example, in atoms, molecules and biological
systems~\cite{haessler, tzallas, hockett, niikura, ihee}.

\begin{acknowledgments}
We thank Jan Malte Slowik for careful reading of the manuscript.
\end{acknowledgments}

\end{document}